# Scatting theory for reflective Fourier ptychographic diffraction tomography

Shuhe Zhang*, Tos T. J. M. Berendschot and Shuo Zhang

University Eye Clinic Maastricht, Maastricht University Medical Center+, P.O. Box 5800, 6202 AZ
Maastricht, the Netherlands

*shuhe.zhang@maastrichtuniversity.nl

A forward model is presented to an inverse scattering problem that arises in the application of reflective Fourier ptychographic microscopy. The model allows us to determine the 3D distributions of refractive index for weakly scattering semi-transparent objects using Fourier ptychographic tomography. The derivation results show that both the transmission type reported previously and the reflective type present in this article are corresponding to a specific application in Wolf's work (Emil Wolf, Opt. Commun. 1969 1(4) pp. 153-156). The transmission type measures the forward scattering field while the reflective type measures the backward scattering field. This model is also available for holographic method.

The transmission type of Fourier ptychographic microscopy (tFPM) shows its attractive applications biomedical researches, where the thin sample is the basic assumption in FPM [1]. Later, researchers have found that the FPM can be also applied to the thick sample, and two physical models (1) the multi-slice model [2, 3] and (2) the Fourier scattering model [4, 5] were developed, extending the FPM to Fourier ptychographic tomography (FPT).

Compared with the success of transmission type, the reflective type of FPM (rFPM) is rarely studied. Only a few researches have been reported the rFPM [6-8], and most of them focused on the modification of the illumination path of rFPM. However, the physical models for forming an image in a reflective type of FPM are lack of discussion even though the reflection is a more complex situation than the transmission, especially for thick samples. Of course, optimizing the illumination path is important for a rFPM due to the special illumination requirement in an FPM system, while a proper physical model for rFPM is also important since it shows that in what situation can an rFPM work properly and produce correct results.

In thin sample model of tFPM, the transmitted field while the thin sample is illuminated by a tilted plane wave is the product of the sample's transmission function and the illumination field, while the rFPM directly employs the same idea in which the reflected field is the product of the sample's reflection function and the illumination field. However, the diffuse reflection of a thick sample may make FPM fail to reconstruct. Therefore, a strengthened physical model for imaging formation of rFPM is needed.

In this research, we focus on the weakly scattering sample and derive the Fourier scattering model rFPM using first Born approximation. This model shows that under some assumptions that can be easily achieved experimentally, it is possible to recover the 3D structure of an object using rFPT. The model also can be reduced to 2D situation corresponding to a typical 2D rFPM.

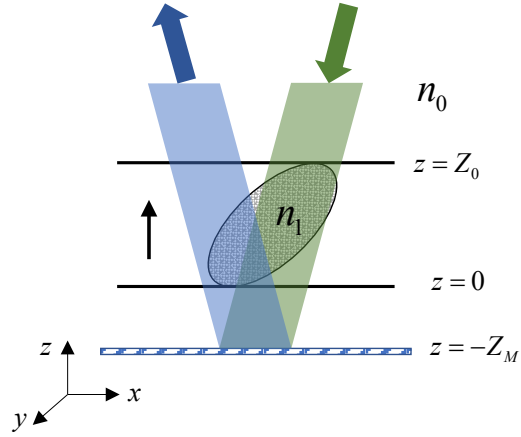

Fig. 1. Illustration of a reflective microscopy.

Let's consider the imaging geometry in Fig. 1. A 3-D object of unknown permittivity and height is distributed in a known media with a refractive index $n_1 > n_0$ within $z \in [0, Z_0]$. Let notation $n = n_1 / n_0$ be the relative refractive index. For simplicity, we assume that the object is situated in free space where $n_0 = 1$. A mirror is placed at $z = -Z_M$ below the object. With this geometry, the total field must consider the reflections of both the object's scattered field and the illumination from the interface. Considering the object is illuminated by a monochromatic plane wave $U_F = \exp(i\mathbf{k}_{in} \cdot \mathbf{r})$ with central wavelength $\lambda$ at arbitrary oblique angles, where $\mathbf{k}_{in} = (k_{x,in}, k_{y,in}, -k_{z,in})$ is the wave vector of the plane wave and $k_{z,in} = \sqrt{k_0^2 - k_{x,in}^2 - k_{y,in}^2}$, $k_0 = 2\pi / \lambda$. $\mathbf{r} = (x, y, z)$ is the spatial coordinates vector. This plane waves can either come from inside of the objective lens [6] or emit from a light source outside the detection path [8]. Due to the existing of the mirror, the total illumination field is the sum of the incident plane wave $U_F$ and

its reflected field $U_B$

$$U_{in} = U_F + U_B = \exp(i\mathbf{k}_{in}\mathbf{r}) + R(\mathbf{k}_{in,\perp})\exp(i\mathbf{k}_{in,\perp}\mathbf{r}_\perp)\exp(ik_{z,in}z)\exp(2ik_{z,in}Z_m), \qquad (1)$$

where $\mathbf{k}_{in,\perp} = (k_{x,in}, k_{y,in})$ and $\mathbf{r}_\perp = (x, y)$. $R(\mathbf{k}_{in,\perp})$ is the average TE and TM wave Fresnel coefficients of the mirror for unpolarized illumination. Here, $U_B = R(\mathbf{k}_{in,\perp})\exp(i\mathbf{k}_{in,\perp}\mathbf{r}_\perp)\exp(ik_{z,in}z)\exp(2ik_{z,in}Z_m)$ and $\exp(2ik_{z,in}Z_m)$ is the accumulated phase of the reflected field $U_B$ when it arrived at $z=0$ plane again.

Let $U_{sc}$ be the scattering field and $U_{tot} = U_{in} + U_{sc}$ be the total field. Our goal is to calculate the unknown scattering field from the object. Since the objective lens can only collect the field propagating in positive z-axis, the field before entering the objective lens is $U_B + U_{sc}$.

We consider a weakly scattering object, following Wolf's work [9], the $U_{sc}$ and $U_{tot}$ obey the scattering wave equation

$$(\nabla^2 + k_0^2)U_{sc}(\mathbf{r}) = F(\mathbf{r})U_{tot}(\mathbf{r}), \qquad (2)$$

where $F(\mathbf{r}) = -k_0^2[n^2(\mathbf{r}) - 1]$ is known as the scattering potential of the object. A good approximation to the solution of Eq. (2) for the scattered field is given by the first Born approximation,

$$U_{sc}(\mathbf{r}) = -\frac{1}{4\pi}\int F(\boldsymbol{\rho})U_{in}(\boldsymbol{\rho})G(\mathbf{r}-\boldsymbol{\rho})\mathrm{d}^3\boldsymbol{\rho}, \qquad (3)$$

where $\boldsymbol{\rho} = (x_\rho, y_\rho, z_\rho)$ is the spatial coordinate of inside the object, and $G(\mathbf{r}-\boldsymbol{\rho})$ is the 3D Green function viz, a diverging spherical wave that propagates from $\boldsymbol{\rho}$ toward $\mathbf{r}$.

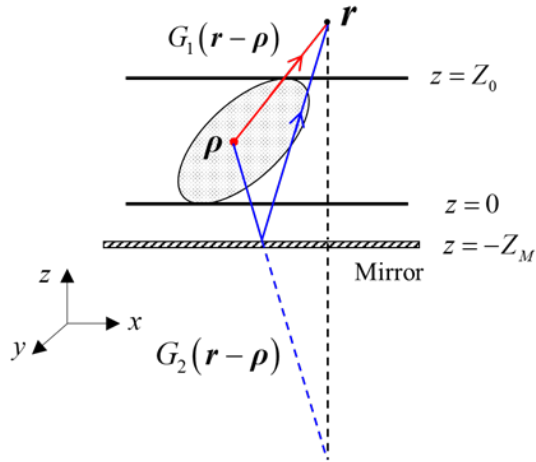

Fig. 2. Separation of Green function.

Due to the mirror reflection, $G(r-\rho)$ can be separated into two components as shown in Fig. 2, the first component is denoted by $G_1(r-\rho)$ which directly propagates from $\rho$ to $r$ (red line), while the second component is denoted by $G_2(r-\rho)$ which first arrived at the mirror and then reflected toward $r$ (blue folded line). Combined with the Weyl expansion of an upper half-space spherical wave, $G_1(r-\rho)$ and $G_2(r-\rho)$ are written as

$$G_1(r-\rho) = \frac{i}{2\pi} \int_{-\infty}^{+\infty}\int_{-\infty}^{+\infty} \frac{1}{k_z} \exp\left\{i\left[k_x(x-x_\rho) + k_x(y-y_\rho) + k_z(z-z_\rho)\right]\right\} dk_x dk_y, \tag{4.1}$$

and

$$G_2(r-\rho) = \frac{i}{2\pi} \int_{-\infty}^{+\infty}\int_{-\infty}^{+\infty} \frac{1}{k_z} R(k_x,k_y) \exp\left\{i\left[k_x(x-x_\rho) + k_y(y-y_\rho) + k_z(z+z_\rho+2Z_m)\right]\right\} dk_x dk_y. \tag{4.2}$$

Here $k_z = \sqrt{k_0^2 - k_x^2 - k_y^2}$. According to Eqs. (3) and (4), $U_{sc}(r)$ can be expressed as two components as well, yielding $U_{sc}(r) = U_{sc,1}(r) + U_{sc,2}(r)$ with $U_{sc1}(r) = -(4\pi)^{-1} \int F(\rho) U_{in}(\rho) G_1(r-\rho) d^3\rho$ and $U_{sc2}(r) = -(4\pi)^{-1} \int F(\rho) U_{in}(\rho) G_2(r-\rho) d^3\rho$.

Let's first consider the $U_{sc1}(r)$. According to Eq. (3) and Eq. (4.1), $U_{sc1}(r)$ is

$$\begin{aligned} U_{sc1}(r) = -\frac{i}{8\pi^2} \int F(\rho) \exp(ik_{x,in}x_\rho + ik_{y,in}y_\rho) \\ \times \left[\exp(-ik_{z,in}z_\rho) + R(k_{x,in},k_{y,in})\exp(ik_{z,in}z_\rho)\exp(2ik_{z,in}Z_m)\right] \\ \times \left\{\int_{-\infty}^{+\infty}\int_{-\infty}^{+\infty} \frac{1}{k_z}\exp\left\{i\left[k_x(x-x_\rho)+k_y(y-y_\rho)+k_z(z-z_\rho)\right]\right\}dk_x dk_y\right\} d^3\rho \end{aligned} \tag{5}$$

Changing the integration order, we obtain

$$\begin{aligned} U_{sc1}(r) = -i\pi \int_{-\infty}^{+\infty}\int_{-\infty}^{+\infty} \frac{1}{k_z} \Big\{ R(k_{x,in},k_{y,in}) \exp(2ik_{z,in}Z_m) \tilde{F}^{(3)}(k_x-k_{x,in}, k_y-k_{y,in}, k_z-k_{z,in}) \\ + \tilde{F}^{(3)}(k_x-k_{x,in}, k_y-k_{y,in}, k_z+k_{z,in})\Big\} \exp\left[i(k_x x+k_y y+k_z z)\right] dk_x dk_y \end{aligned}, \tag{6}$$

where

$$\tilde{F}^{(3)}(k_x,k_y,k_z) = \frac{1}{(2\pi)^3} \int F(\rho)\exp\left[-i(k_x x_\rho + k_y y_\rho + k_z z_\rho)\right] d^3\rho \tag{7}$$

is the 3D Fourier transform of $F(\rho)$ with respect to spatial coordinate $\rho$. Now fixing $U_{sc1}(r)$ at a given plane $z$, and performing 2D Fourier transform with respect to $x$ and $y$ to both sides of Eq. (6), we obtain

$$\tilde{U}_{sc1}^{(2)}(k_x, k_y, z) = -\frac{i}{4\pi} \frac{\exp(ik_z z)}{k_z} \Big[ \tilde{F}^{(3)}(k_x - k_{x,in}, k_y - k_{y,in}, k_z + k_{z,in}) \\ + R(k_{x,in}, k_{y,in}) \tilde{F}^{(3)}(k_x - k_{x,in}, k_y - k_{y,in}, k_z - k_{z,in}) \exp(2ik_{z,in} Z_m) \Big] \quad (8.1)$$

where

$$\tilde{U}_{sc1}^{(2)}(k_x, k_y, z) = \frac{1}{(2\pi)^2} \int U_{sc1}(\mathbf{r}_\perp, z) \exp\left[-i(k_x x + k_y y)\right] d^2 \mathbf{r}_\perp \quad (8.2)$$

is the 2D Fourier transform of $U_{sc1}(\mathbf{r}_\perp, z)$. Similar treatment can be also applied to $U_{sc2}(\mathbf{r}_\perp, z)$, resulting in

$$\tilde{U}_{sc2}^{(2)}(k_x, k_y, z) = -\frac{i}{4\pi} \frac{\exp[ik_z(z + 2Z_m)]}{k_z} \Big[ R(k_x, k_y) \tilde{F}^{(3)}(k_x - k_{x,in}, k_y - k_{y,in}, -k_z + k_{z,in}) \\ + R(k_{x,in}, k_{y,in}) R(k_x, k_y) \tilde{F}^{(3)}(k_x - k_{x,in}, k_y - k_{y,in}, -k_z - k_{z,in}) \exp(2ik_{z,in} Z_m) \Big] \quad (9)$$

Combined with Eq. (8) and Eq. (10), the 2D Fourier transform of $U_{sc}(\mathbf{r})$ is given by

$$\tilde{U}_{sc}^{(2)}(\mathbf{k}) = -\frac{i}{4\pi} \frac{\exp(ik_z z)}{k_z} \left[ FF(\mathbf{k}) + FB(\mathbf{k}) + BF(\mathbf{k}) + BB(\mathbf{k}) \right], \quad (10)$$

where

$$\begin{cases} FF(\mathbf{k}) = \tilde{F}^{(3)}(k_x - k_{x,in}, k_y - k_{y,in}, -k_z + k_{z,in}) R(k_x, k_y) \exp(2ik_z Z_m) \\ FB(\mathbf{k}) = \tilde{F}^{(3)}(k_x - k_{x,in}, k_y - k_{y,in}, k_z + k_{z,in}) \\ BF(\mathbf{k}) = \tilde{F}^{(3)}(k_x - k_{x,in}, k_y - k_{y,in}, k_z + k_{z,in}) R(k_{x,in}, k_{y,in}) \exp(2ik_{z,in} Z_m) \\ BB(\mathbf{k}) = \tilde{F}^{(3)}(k_x - k_{x,in}, k_y - k_{y,in}, -k_z - k_{z,in}) R(k_x, k_y) R(k_{x,in}, k_{y,in}) \exp(2ik_{z,in} Z_m) \exp(2ik_z Z_m) \end{cases} \quad (11)$$

Components *FF* and *FB* denote the forward and backward scattering field when the object is illuminated by $U_F$. *FF* is reflected by the mirror and arrived at *r*, while *FB* propagates to *r* directly. Components *BF* and *BB* denote the forward and backward scattering field when the object is illuminated by $U_B$ which is the reflected field of $U_F$, hence an additional term $R(k_{x,in}, k_{y,in})$ exists for both *BF* and *BB*. Under the illumination of $U_B$, the forward scattering component *BF* directly propagates to *r*, while *BB* is reflected by the mirror. When the object is attached to the mirror, we have $Z_m = 0$.

The total field at the entrance of the objective lens is $U_B + U_{sc}$, and the low-passed by the pupil function to form an image at the camera sensor. The camera sensor measuring only the intensity of the field,

$$I = \left| \mathcal{F}^{-1} \left[ \mathcal{F}(U_B + U_{sc}) P(\mathbf{k}) \right] \right|^2, \tag{12}$$

where $P(\mathbf{k})$ is the pupil function of the objective lens. It has been provided experimentally that the complex amplitude of $U_{sc}$ can be recovered through the Fourier ptychographic diffraction tomography in a transmission system [4, 5].

Different from transmission type [4, 5, 9], Eq. (12) is nonlinear and cannot be used directly for determination of $\tilde{F}^{(3)}(k_x, k_y, k_z)$ by measuring $U_{sc}$, we therefore need additional assumptions on the field behavior to maintain a simplified nonlinear physical model.

**Special case 1: $R(k_x, k_y) = 1$ and thin object at z = 0.**

In this case, $FF = FB = BF = BB$, Eq. (11) is then reduce to a linear model, and the refractive index of the thin object can be recovered by a normal reflective Fourier ptychographic microscopy [7, 8, 10].

**Special case 2: $R(k_x, k_y) \ll 1$ and thin object at z = 0.**

In this case where $R(k_x, k_y) \ll 1$, the mirror can be regarded as either being removed, or a dark surface. The only contribution to the scattering field is the back-scattering under the illumination of $U_F$. $\tilde{U}_{sc}^{(2)}(\mathbf{k}) = -i \exp(ik_z z) FB(\mathbf{k}) / (4\pi k_z)$, which is linear. Practically, this case does exist, for example when we are observing a single positive USAF target in a reflective microscopy. The background of the USAF target is almost dark, while the stripes can be clearly observed. Moreover, since $R(k_x, k_y) \ll 1$, the intensity contribution of reflected illumination field $U_B$ can also be ignored thus the dark-field-like images are observed.

**Special case 3: $R(k_x, k_y) \ll 1$ and thick object.**

The part of the 3D information of the $FB(\mathbf{k})$ can be recovered by measuring the intensity of $U_{sc}$ as in a transmission type of Fourier ptychographic tomography. Originally, in Wolf's research [9], it is proposed that one can measure the forward scattering field and backward scattering field at both sides of the object. Both forward scattering field and backward scattering field determine parts of the 3D Fourier spectrum of the scattering potential. While in a transmission type of Fourier ptychographic tomography is corresponding to the forward measurement of the scattering field, while a reflective type is corresponding to the backward measurement of the scattering field, in the case of $R(k_x,k_y)$. The 3D coherent transfer function (CTF) is therefore similar to the transmission type but flip up and down as shown in Fig. 3.

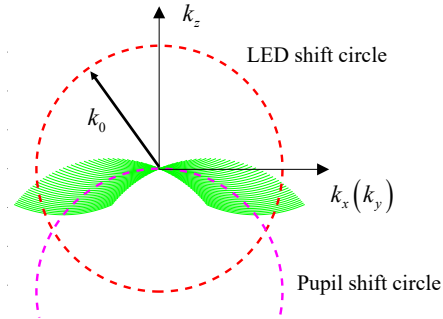

Fig. 3. 3D CTF of rFPT.